\documentclass[prl,aps,epsf,twocolumn,showpacs,10pt,superscriptaddress ]{revtex4-1}

\usepackage{graphicx,amsfonts,times,bm,amsmath,verbatim,color,array}

\newcommand{\e}{\varepsilon}

\usepackage{natbib}
\usepackage{graphicx,amssymb,amsmath,ifthen,braket,xcolor}
\usepackage{bm}

\newcommand{\angstrom}{\mbox{\normalfont\AA}}

\begin{document}


\title{Fundamental relations for anomalous thermoelectric transport coefficients in the non-linear regime}

\author{Chuanchang Zeng}
\affiliation{Department of Physics and Astronomy, Clemson University, Clemson, SC 29634, USA}
\author{Snehasish Nandy}
\affiliation{Department of Physics, University of Virginia, Charlottesville, VA 22904, USA}
\author{Sumanta Tewari}
\affiliation{Department of Physics and Astronomy, Clemson University, Clemson, SC 29634, USA}


\begin{abstract}
In a series of recent papers anomalous Hall and Nernst effects have been theoretically discussed in the non-linear regime and have seen some early success in experiments.
In this paper, by utilizing the role of Berry curvature dipole, we derive the fundamental mathematical relations between the anomalous electric and thermoelectric transport coefficients in the non-linear regime. The formulae we derive replace the celebrated Wiedemann-Franz law and Mott relation of anomalous thermoelectric transport coefficients defined in the linear response regime. In addition to fundamental and testable new formulae, an important byproduct of this work is the prediction of nonlinear anomalous thermal Hall effect which can be observed in experiments.
\end{abstract}

\maketitle

{\color{blue}{\em Introduction}}---Onsager's reciprocity relations mandate that the Hall effect in linear response has to vanish in a time reversal invariant system whereas the non-linear Hall effect has no such restriction \cite{LD_book}. The generalized Onsager's relation appropriate for non-linear current response indicates that in order to get a non-zero DC non-linear conductivity, the current response requires dissipation and should be proportional to the relaxation time $\tau$ \cite{Naoto_SP_2018}. Unlike the anomalous Hall effect in the linear-response regime \cite{Naoto_2010review_berryphase,Niu_2010_berryphase,Ye_1999_berryphase,Tokura2001_AHE,Niu2002_AHE,Qian_2003_berryphase,Nagaosa2003_AHE,Lee2004_AHE,Fang2003_AHE, Haldane2004_AHE,cxliu_2008_breakingTRS,Yu2010_AHE,qikun2013_AHE}, the non-linear anomalous Hall effect (NLAHE) does not require broken time reversal symmetry (TRS) but needs inversion symmetry (IS) breaking. The Berry curvature dipole (BCD), which is defined as the first order moment of the Berry curvature over the occupied states, is found to be responsible for NLAHE \cite{JMoore2010_Berryphase_factor,Inti_2015_BCD, kKang2019_NLAHE_experiment,qMa2019_NLAHE_experiment,Nandy_sysmetris_2019,Law2019_NLAHE_3}. 
The importance of electron-electron interactions for the external magnetic field dependence of the non-linear conductivities have been pointed out \cite{eeinteraction_1,eeinteraction_2}.
Motivated by the idea of NLAHE, another second-order response function, non-linear anomalous Nernst effect (NLANE), has been predicted in transition metal dichalcogenides (TMDCs) \cite{Naoto_NLANE_3,Yu_NLANE_1,Zeng_NLANE_2}. Interestingly, these nonlinear responses could manifest distinctive behaviors and have become promising tools for understanding novel materials with low crystalline symmetry in experiments. In this paper, by utilizing the role of the Berry curvature dipole, we derive the fundamental mathematical formulae among the anomalous electric and thermoelectric transport coefficients in the non-linear regime, replacing the celebrated Wiedemann-Franz law and the Mott relation \cite{Ashcroft} which are valid in the linear response regime.

In this paper, we begin with the derivation of a new non-linear response function, namely, the non-linear anomalous thermal Hall effect (NLATHE), which can be directly observed in experiments. NLATHE refers to the appearance of a transverse thermal gradient as a second order response to an applied longitudinal heat current (Fig.~1).
\begin{figure}[b]
\vspace{-6mm}
	\begin{center}
		\includegraphics[width=0.35\textwidth]{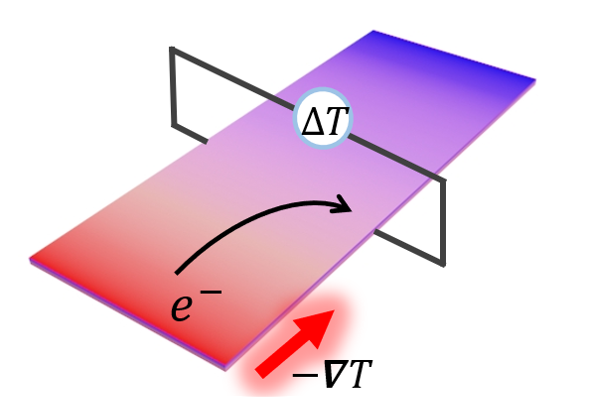}
	\end{center}
	\caption{(Color online) Schematic experimental setup for measuring the non-linear anomalous thermal Hall effect. A transverse thermal gradient ($\Delta T$) can be measured as a second-order response of the longitudinal heat current even in the absence of external magnetic field. The sample breaks inversion but respects time reversal symmetry, so the linear response anomalous thermal Hall current is known to vanish by symmetry. }\label{fig:setup}
\end{figure}
Armed with these calculations we then address the question of fundamental relations among the anomalous transport coefficients in the non-linear regime. In linear response theory, the relations among electric, thermo-electric and thermal transport coefficients of metals are encapsulated by the celebrated Wiedemann-Franz law and Mott formula \cite{Ashcroft}.  These formulae in the context of linear anomalous transport coefficients have been studied in topologically trivial and non-trivial materials in theory as well as experiments \cite{Xiao_2006_THE_wavepacket,QIN_2011_THE_wavepacket,lifa_2016_THE_wavepacket,Niu_2010_berryphase, Jens_2017_THE,Lundgren_2014_THE,Yokoyama_2011_THE_heat,Girish_2016_THE,Timothy_2017_THE, Xiao_2006_THE_wavepacket,QIN_2011_THE_wavepacket}. While according to the Wiedemann-Franz law, the electric and thermal conductivities (regular or anomalous) are directly proportional to each other, the Mott formula predicts that the Nernst coefficient is proportional to the derivative of the Hall coefficient with respect to the chemical potential (see Eq.~(7, 8)).  Interestingly, our analytical calculations for all three anomalous transport coefficients allow us to  predict fundamentally new relations among the transport coefficients in the non-linear regime. The principal result of this work is the remarkable prediction that in the non-linear regime the anomalous Hall and Nernst coefficients are directly proportional to each other (Eq.~(16)), while they are related through a derivative in the linear response regime (Mott relation, Eq. (8)). Moreover, the derivative appears in the formula relating the electric and thermal conductivities in the non-linear regime (Eq. (14)), while the Wiedemann-Franz law (Eq.~(7)) in the linear response regime has no such derivative. The role of the derivative is thus interchanged in the non-linear regime with respect to its linear response counterpart. These results should be tested in experiments as confirmation of the intrinsic non-linearity, rather than a more conventional departure from the Wiedemann-Franz law and the Mott formula. We check the validity of our analytical results by full numerical evaluation of the relevant quantities for MoS$_2$, a TR invariant but inversion symmetry broken TMDC that has been intensively studied in experiments recently.

{\color{blue}{\em Boltzmann theory and anomalous thermal Hall effect in non-linear regime}}---The phenomenological Boltzmann transport equation can be written as
\begin{equation}
    \{ \partial_t + {\dot{\bm{r}} \bm{\nabla_r}+ \dot{\bm{k}} \bm{\nabla_k}} \}f(\bm{k,r},t)  = I_{coll}\{f(\bm{k,r},t)\}
\label{e1}
\end{equation}
where the collision integral $I_{coll}\{f(\bm{k,r},t)\}$ incorporates the effects of electron correlations (inelastic scattering) and elastic scattering from
impurities. For the sake of simplicity, we here focus only on the impurity scattering. Invoking the relaxation time approximation, the steady-state solutions to the Boltzmann equation is given by
\begin{equation}
    \begin{split}
      \{ {\dot{\bm{r}} \bm{\nabla_r}+ \dot{\bm{k}} \bm{\nabla_k}} \}f(\bm{k})=-\frac{g_{\bm{k}}}{\tau}
    \end{split} \label{eq:collision_term}
\end{equation}
where $g_{\bm{k}}=f(\bm{k})-f_0$ is the difference between the perturbed Fermi-Dirac distribution $f_{\bm{k}}$ and equilibrium Fermi-Dirac function $f_0$. Considering the homogeneous uniform fields, we have dropped the $\bm{r}$ dependence of $f(\bm{k,r},t)$. Here, $\tau$ is the average scattering time between two successive collisions.  For simplicity we ignore the momentum dependence of the scattering time $\tau$ and assume it to be a constant for this work.

To find the non-linear anomalous thermal Hall coefficient in the absence of the external fields, we expand $g_{\bm{k}}$ as $g_{\bm{k}}=g^1_{\bm{k}}+g^2_{\bm{k}}+ ...$, where $g^n_{\bm{k}}$ is understood as the $n^{th}$ order response to the applied thermal gradient, i.e., $g^n_{\bm{k}} \propto (\bm{\nabla}T)^n$. Substituting $f({\bm{k}})=f_0 +g_{\bm{k}}$ into the steady-state Boltzmann equation given in Eq.~(\ref{eq:collision_term}), we could find the distribution function at the first and second-order in the thermal gradient as,
\begin{equation}
    \begin{split}
        g^1_{\bm{k}} & = \tau \bm{v_k} \frac{(\e_{\bm{k}}-\mu)}{T}\frac{\partial f_0 }{\partial \e_{\bm{k}} } \bm{\nabla}T, \\
        g^2_{\bm{k}} & = \tau \bm{v_k} \frac{(\e_{\bm{k}}-\mu)}{T}\frac{\partial g^1_{\bm{k}} }{\partial \e_{\bm{k}} } \bm{\nabla}T, \\
        \end{split} \label{eq:perturbed_distributions}
\end{equation}
where $\mu$ is the chemical potential, $\bm{v_k}=\hbar^{-1} \bm{\nabla _k} \e_{\bm{k}}$ is group velocity with $\e_{\bm{k}}$ the energy dispersion. 
In principle, expansions with higher orders ($\nabla T$)  around the equilibrium distribution function can be derived by iteration. However, in this paper we restrict the expansions only up to the quadratic order and neglect the small higher order ($\propto \mathcal{O}(\tau^n)$, $n\geq 3$) terms.
\begin{figure}[htp!]
	\begin{center}
		\includegraphics[width=0.46\textwidth]{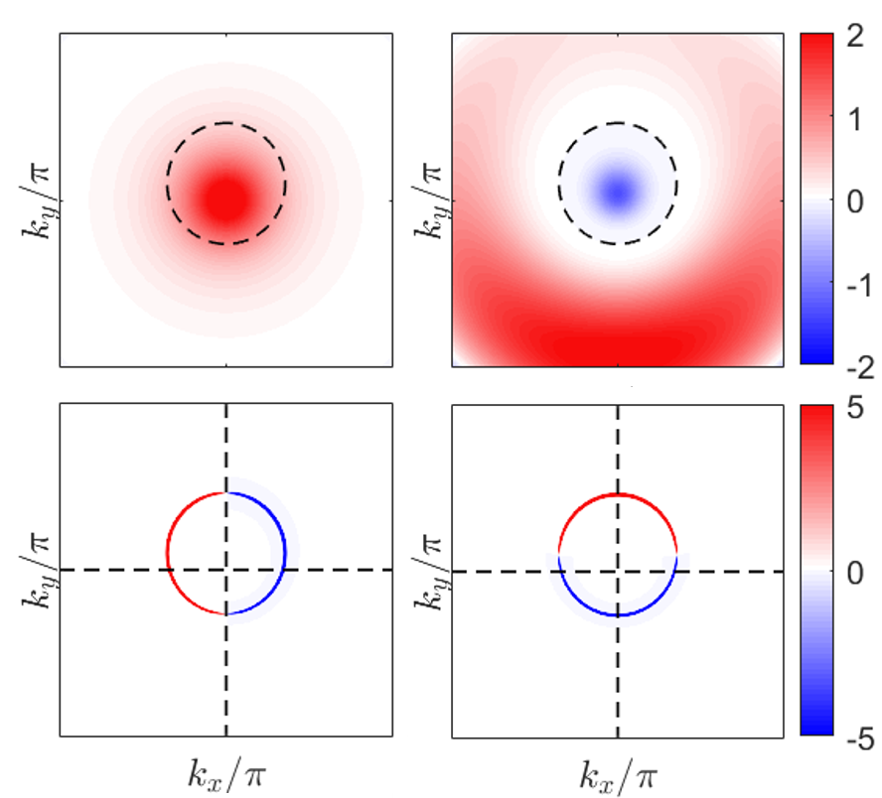}
		\llap{\parbox[b]{150mm}{\large\textbf{(a)}\\\rule{0ex}{66mm}}}
		\llap{\parbox[b]{78mm}{\large\textbf{(b)}\\\rule{0ex}{66mm}}}
		\llap{\parbox[b]{153mm}{\large\textbf{(c)}\\\rule{0ex}{31mm}}}
		\llap{\parbox[b]{82mm}{\large\textbf{(d)}\\\rule{0ex}{31mm}}}
	\end{center}
	\vspace{-5mm}
	\caption{(Color online) (a) Berry curvature $\bm{\Omega_k}^{n,s}$ and (b) modulated Berry curvature $ \beta^3 (\mathcal{E}_{\bm{k}}-\mu)^3 \bm{\Omega_k}^{n,s}$ projected on the $\bm{k}$ space for Hamiltonian given in Eq.~(\ref{eq:generalHamiltonian}). The black dash lines indicate the Fermi surface at $\mu=1.5\Delta$. Panel (c) and (d) show the derivative of Fermi distribution function at Fermi energy $\mu=1.5\Delta$ for $\partial_x f_{\bm{k}}$ and $\partial_y f_{\bm{k}}$ respectively. The parameters used here are $n=1, s=1, t=1.1 eV, a=3.19 \angstrom, v=a t, \alpha= 0.1 v, \Delta=1.8 eV$, $k_{x,y} \in [-0.5\pi,0.5\pi]$, $\beta=1 (eV)^{-1}$ is considered for (a), (b) and temperature $T=100K$ is applied for (c), (d).
	} \label{fig:k_plots}
\end{figure}

After accounting for both the normal and anomalous contributions, the total thermal current $\bm{j}^Q_{tot}$ is given by $\bm{j}^Q_{tot} =\bm{j}^Q_N +\bm{j}^Q_E +\bm{j}^Q_T $, where $\bm{j}^Q_N$ is the standard contribution to thermal current coming from the conventional velocity $\bm{v_k}$ of the carriers, and $\bm{j}^Q_E$ is the anomalous thermal current mediated by the Berry curvature $\bm{\Omega_k}$ in the presence of electric field $\bm{E}$ \cite{Xiao_2006_THE_wavepacket}. In this paper, we are interested in the last term $\bm{j}^Q_T$ given by \cite{Doron_2010},
\begin{equation}
    \begin{split}
        \bm{j}^Q_T = & - \frac{k^2_B T}{\hbar} \bm{\nabla}T \times \int [d\bm{k}] \sum_n \bm{\Omega_{\bm{k}}}^n  \bigg[  \beta^2 \left(\e^n_{\bm{k}}-\mu \right)^2 f_0 \\ &+ \frac{\pi^2}{3} -\operatorname{In}^2\left(1-f_0\right)- 2\operatorname{Li}_2\left(1-f_0\right)\bigg]
    \end{split} \label{eq:THE_linear}
\end{equation}
which describes the transverse thermal response to the applied thermal gradient $-\bm{\nabla}T$ in the presence of a non-trivial Berry curvature $\bm{\Omega_k}$.

Substituting Eq.~(\ref{eq:perturbed_distributions}) into the thermal Hall term in Eq.~(\ref{eq:THE_linear}) (with $f_0$ replaced by $f_{\bm{k}}=f_0 +g_{\bm{k}}$)), the non-linear anomalous thermal Hall current flowing along the direction $a$ (second order of $-\bm{\nabla}T$) can be written as
\begin{equation}
    \begin{split}
    \left(\bm{j}^Q_T\right)^{\prime}_a=\epsilon_{abc}\frac{\tau\nabla_{b}T\nabla_{d}T}{\hbar^2} \int [d\bm{k}] \sum_n \Omega^n_{\bm{k},c} \frac{\left(\e^n_{\bm{k}}-\mu \right)^3}{T^2}  \frac{\partial f_0}{\partial k_d}
    \end{split}
\end{equation}
where the prime on $(\bm{j}^Q_T)^{\prime}$ indicates the nonlinear response and $a,b,c,d$ represent the components $x,y,z$, and $n$ is the band index.
In this paper we focus on this Berry curvature-dependent anomalous contribution to  $(j^Q_T)^{\prime}_a$ which is non-zero in TRS invariant systems. 

From Eq.~(5), the non-linear anomalous thermal Hall coefficient can be written as $\big[(j^Q_T)^{\prime}_a = \epsilon_{abc}l^{\prime}_{cd}  (\nabla_{b}T\nabla_d T)\big]$,
\begin{equation}
    \begin{split}
    l^{\prime}_{cd} =\frac{\tau T}{\hbar^2} \int [d\bm{k}] \sum_{n} \Omega^{n}_{\bm{k},c} \frac{\left(\e^{n}_{\bm{k}}-\mu \right)^3}{T^3}\frac{\partial f_0}{\partial k_d}
    \end{split} \label{eq:coefficient_NLATH}
\end{equation}
This is one of the main results of this paper. We find that NLATHE, which is linearly proportional to the scattering time, appears due to the Berry curvature from the states near the Fermi surface. Under TR symmetry, we know $\bm{\Omega}_{\bm{k}}=-\bm{\Omega}_{\bm{-k}}$, $\e_{\bm{k}}=\e_{-\bm{k}}$ and $\partial f_0 /\partial {k}_d = - \partial f_0 /\partial \left(-{k}_d\right)$. Therefore, it is clear from the Eq.~(6) that NLATHE can survive even in the time-reversal invariant systems. 




{\color{blue}{\em Analog of Wiedemann-Franz law and Mott relation in the non-linear regime}}---We now investigate the celebrated Wiedemann-Franz law and Mott relation in the non-linear regime at low temperatures. In the linear response regime, the Wiedemann-Franz law which gives the ratio between thermal conductivity ($\kappa_{ab}$) and electrical conductivity ($\sigma_{ab}$), is given by \cite{Lorenz_law}
\begin{equation}
    \begin{split}
       \frac{ \kappa_{ab} }{\sigma_{ab}}& =L T
    \end{split}  \label{eq:L_wfLaw}
\end{equation}
with $L=\pi^2 k_B^2/3e^2$ the Lorentz number.
On the other hand, the Mott relation can be written as \cite{Ashcroft}
\begin{equation}
    \alpha_{ab}= e L T \frac{\partial \sigma_{ab}}{\partial \mu}.
 \label{eq:Motts}
\end{equation}
where $\alpha_{ab}$ is the thermo-electric conductivity. To derive the analog of these formulas in the non-linear regime, we first consider the non-linear anomalous Hall effect. The BCD induced NLAHE at a finite temperature can be written as ~\cite{Inti_2015_BCD},
\begin{equation}
    \begin{split}
        \chi_{abc} = \epsilon_{abd}\frac{e^3 \tau}{2\hbar^2} D_{cd}
    \end{split} \label{eq:coefficient_NLAHE}
\end{equation}
where $D_{cd}$, the Berry curvature dipole, is defined as
\begin{equation}
    \begin{split}
      D_{cd}=\sum_n \int [d\bm{k}]  \frac{\partial \bm{\Omega}_{\bm{k},c}} {\partial k_d} f_{\bm{k}} = -\sum_n \int [d\bm{k}] \bm{\Omega}_{\bm{k},c}  \frac{\partial f_{\bm{k}}}{\partial k_d}
    \end{split}
\end{equation}
Using the Sommerfeld expansion \cite{Ashcroft}, the BCD term ($D_{cd}$) of NLAHE at low temperature can be written as
\begin{equation}
    D_{cd}(T,\mu) = G_{cd}(\mu)+\frac{\pi^2}{6}(k_B T)^2 G^{(2)}_{cd}(\mu)+ \mathcal{O}(T^4) \label{eq:NLAHE_Tdependence}
\end{equation}
where
\begin{equation}
    \begin{split}
        G_{cd}(\e) &= \int [d\bm{k}] \delta(\e-\e_k)\bm{\Omega}_{\bm{k},c} \frac{\partial \e_{\bm{k}}}{\partial k_d}
    \end{split}
\end{equation}
and $G^{(n)}_{cd}(\mu)=\partial^n G_{cd}(\mu) /\partial \mu^n$. Here, the first term $G_{cd}(\mu)$ is the zero-temperature BCD at Fermi energy $\mu$ whereas the second term shows a $T^2$ temperature dependence of the NLAHE which agrees well with previous experimental results \cite{kKang2019_NLAHE_experiment}. Similarly, the NLATHE at low temperature can be written as
\begin{equation}
    l^{\prime}_{cd}(T,\mu) = -\frac{7 \tau \pi^4 k^4_B}{15 \hbar^2} T^2 G^{(1)}_{cd}(\mu) + \mathcal{O}(T^4)  \label{eq:NLTHE_Tdependence}
\end{equation}
with the higher order derivatives $G^{(n)}_{cd}(\mu)$ (odd number $n \geq 3$) included in $\mathcal{O}(T^4)$.

Now, based on Eqs.~(\ref{eq:NLAHE_Tdependence}) and (\ref{eq:NLTHE_Tdependence}), we can write the Wiedemann-Franz law in non-linear regime as
\begin{equation}
    \begin{split}
        l_{cd}^{\prime} & =-\frac{14}{15}e L_0 ^2 T^2 \frac{\partial \chi_{0}(\mu)}{\partial \mu}
    \end{split} \label{eq:NL_wfLaw}
\end{equation}
where $\chi_0(\mu)=e^3 \tau G_{cd}(\mu)/2\hbar^2$ denotes the zero temperature NLAHE coefficient given by Eq.~(\ref{eq:coefficient_NLAHE}), and $L_0 =k^2_B \pi^2/e^2$. Clearly, unlike the linear response regime, where the thermal Hall coefficient and charge Hall coefficient are directly propotional to each other (see Eq.~(7)), the analog of the Wiedemann-Franz law in the non-linear regime given by Eq.~(14) shows that the anomalous thermal Hall coefficient is proportional to the \textit{first order derivative} of the anomalous Hall coefficient with respect to the chemical potential. Also, in contrast to the linear regime, the proportionality factor depends on $T^2$, rather then $T$ as in conventional Wiedemann-Franz law.
The results in Eq.~(\ref{eq:NL_wfLaw}) should be taken as a result of the intrinsic non-linearity, rather than a conventional departure from the Wiedemann-Franz law \cite{Rosa_2013_wfLaw_violation_NLTHE,violation_WF_kim, Rosa_2014_wfLaw_violation,Xu_wfLaw_2018_violation,Jaoui_2018_wfLaw_violation,Yuval_2019_wfLaw_wiolation,Nandy_2019_violation}.

We could also derive the analog of the Mott formula in the non-linear regime by first writing down the NLANE coefficient \cite{Yu_NLANE_1,Zeng_NLANE_2} as,
\begin{equation}
    \begin{split}
        \alpha^{\prime}_{cd}(T,\mu)= & \frac{e\tau}{\hbar^2}\Big\{\frac{\pi^2 k^2_B }{3}G_{cd}(\mu)+\frac{7\pi^4 k_B^4 }{60}T^2 G^{(2)}_{cd}(\mu)+ \mathcal{O}(T^4)\Big\}
    \end{split} \label{eq:NLANE_Tdependence}
\end{equation}
Based on Eqs.~(\ref{eq:NLAHE_Tdependence}) and (\ref{eq:NLANE_Tdependence}), the relation between the coefficients of NLANE and NLAHE can be written as
\begin{equation}
    \alpha^{\prime}_{cd} (\mu)=\frac{2}{3}L_0 \chi_{0}(\mu).
    \label{eq:Motts_NL}
\end{equation}
where we have considered only the first term for $\alpha^{\prime}_{cd}$ (where the higher-order terms are smaller by the successive higher-order directive of the zero temperature BCD, see Eq.~(11)). Eq.~(\ref{eq:Motts_NL}) is the Mott relation in the non-linear regime which shows a finite value for the NLANE ($\alpha^{\prime}$) even at zero temperature. In contrast to the linear regime, where the Nernst coefficient is proportional to the \textit{derivative} of the Hall coefficient (see Eq.~(8)), in the non-linear regime, the corresponding anomalous coefficients are directly proportional to each other. Therefore, we find that, remarkably, the intrinsic non-linearity introduces a derivative in the Wiedemann-Franz law while it removes the same from the Mott relation. These formulas can be directly tested in experiments in TR invariant but inversion broken systems where the anomalous coefficients are zero in the linear regime by symmetry.

{\color{blue}{\em Non-linear transport coefficients for 2D massive Dirac fermions}}---We consider a model Hamiltonian of tilted 2D Dirac cones \cite{Sajedeh_2017_TMDCs,Kin_2016_TMDCs}, which captures the low energy properties of various Dirac materials, such as the surface of topological crystalline insulators and strained transition metal dichalcogenides. The corresponding model Hamiltonian can be written as
\begin{equation}
    \begin{split}
        H_s &= s \alpha k_y \tau_0 + v_F \hbar (k_x \tau_y -s k_y \tau_x) +\Delta \tau_z,\\
    \end{split} \label{eq:generalHamiltonian}
\end{equation}
Here, $v_F$ is the Fermi velocity, $\Delta$ is the energy band gap opened at the $\pm \bm{K}$-valley, $\alpha$ is the tilting parameter and $\tau_{x,y,z,0}$ represent Pauli matrices. The wave vector $\bm{k}$ is measured from the valley center $\pm \bm{K}$ with index $s= \pm 1$ (which also indicates the opposite chirality of the Dirac fermions). Note that the Hamiltonian in Eq.~(17) is TR invariant and the two massive Dirac cones $H_{s=\pm1}$ are mapped to each other by the TR symmetry.
\begin{figure}[t]
	\begin{center}
		\includegraphics[width=0.46\textwidth]{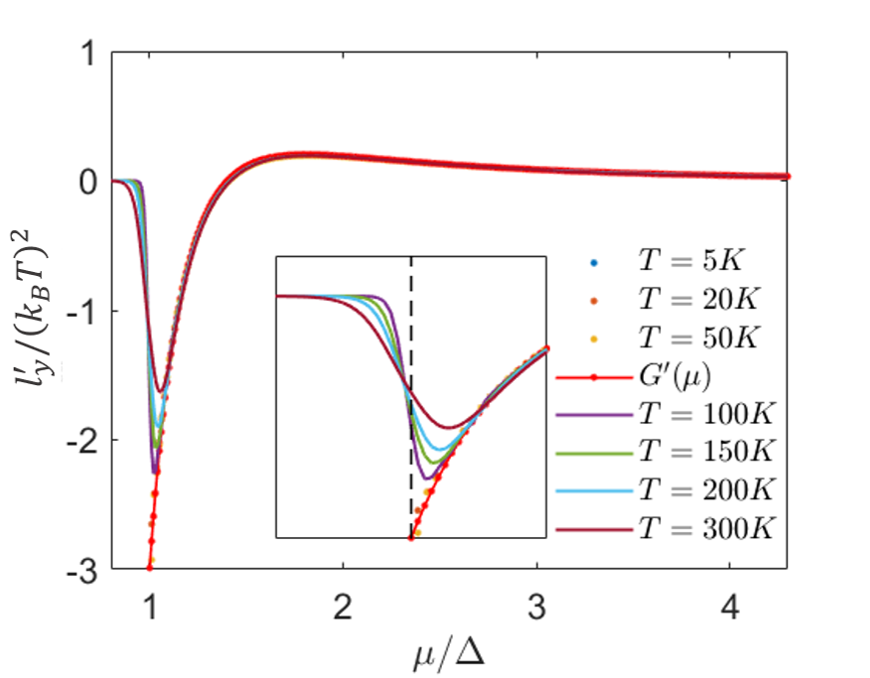}
	\end{center}
    \vspace{-5mm}
\caption{(Color online) Non-linear anomalous thermal Hall coefficient $l^{\prime}_{y}/(k_BT)^2$ versus chemical potential $\mu$ at different temperatures $T$. The red dotted line represents the analytical results based on the non-linear Wiedemann-Franz law (Eq.~(13, 14)). The rest of the data points are results of numerical calculations based on Eq.~(\ref{eq:coefficient_NLATH}).
The inset is a zoom-in of the plot around $\mu=\Delta$ (black dash line). For $\mu< \Delta$ the numerical results deviate from the modified Wiedemann-Franz law valid in the non-linear regime because of the absence of higher orders temperature contributions (see Eq.~(13)). Here the unit for the $y$-axes is $ \tau k^2_B/\hbar^2$, the other parameters are the same as in Fig.~(\ref{fig:k_plots}). } \label{fig:mu_dependence}
\end{figure}

The low energy dispersion and the corresponding Berry curvature of the Hamiltonian are given as,
\begin{equation}
    \begin{split}
        \mathcal{E}^{n,s}_{\bm{k}} &=s\alpha k_y + (-1)^{n-1} \sqrt{\Delta^2+ (v_F \hbar)^2 \bm{k}^2}, \\ \Omega^{n,s}_{\bm{k}} &= (-1)^{n-1} \frac{ s (v_F \hbar)^2 \Delta}{2 (\Delta^2+ (v_F \hbar)^2 \bm{k}^2)^{3/2}}
    \end{split} \label{eq:electric_structures}
\end{equation}
It is clear that $\bm{\Omega_{k}}=-\bm{\Omega_{-k}}$ is satisfied for $\bm{\Omega_{k}}$ in Eq.~(\ref{eq:electric_structures}). The tilting parameter $\alpha$ is required to produce a non-zero Berry curvature dipole contribution which can produce NLAHE, NLANE and NLATHE. In what follows, we use parameters relevant to MoS$_2$, a TR invariant TMDC, to compute the anomalous transport coefficients.

For a system tilted along the $k_y$-axis, only the $x$-direction mirror symmetry ($ \mathcal{M}_x$) that takes $k_x \rightarrow -k_x$ is preserved. As shown as in Fig.~2(a), the Berry curvature $\bm{\Omega_k}$ is azimuthally symmetric in $k_x$-$k_y$ plane,  whereas in Fig.~2(b) the modulated Berry curvature $\beta^3 (\mathcal{E}_{\bm{k}}-\mu)^3 \bm{\Omega_k}$ is only symmetric with respect to $k_x$. Due to the shift of the Fermi surface (black dash line in (a), (b) or the ring in (c), (d)) along $k_y$, the net integral of $\bm{\Omega_k}$ in 2(a) and $\beta^3 (\mathcal{E}_{\bm{k}}-\mu)^3 \bm{\Omega_k}$ in 2(b) over the Fermi surface are non-zero. This explicitly renders the NLATHE a Fermi surface property. Fig.~(\ref{fig:k_plots}) shows that the only non-zero component for NLATHE given in Eq.~(\ref{eq:coefficient_NLATH}) is $l^{\prime}_{zy}$ where $c=z, d=y$ represent $\bm{\Omega}_{\bm{k},z}, \partial_y f_0$ respectively.
\begin{figure}[t]
	\begin{center}
		\includegraphics[width=0.46\textwidth]{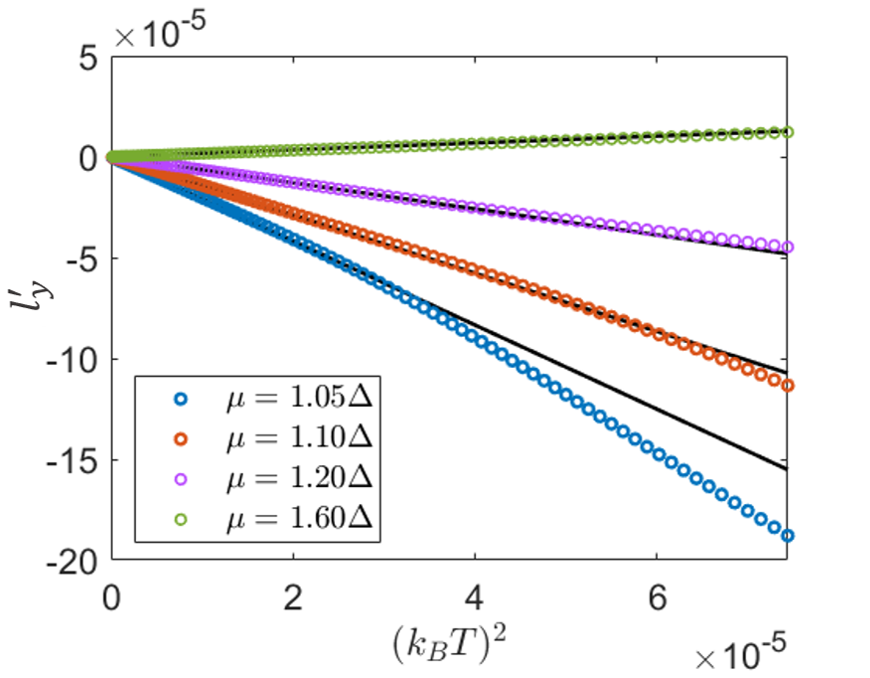}
	\end{center}
	\vspace{-5mm}
	\caption{(Color online) Non-linear thermal Hall coefficient $l^{\prime}_{y}$ plotted as a function of $(k_BT)^2$ for different values of the chemical potential. The circles are from numerical calculations based on Eq.~(6), while the black lines corresponding to each chemical potential are the analytical results based on Eq.~(13, 14). Here the units for $y$-axes is $\tau k^2_B/\hbar^2$, the applied temperature $T\in[5K,100K]$ with a unit step ($1K$), and all other parameters are the same as in Fig.~(\ref{fig:k_plots}). } \label{fig:T_dependence}
\end{figure}

It has been shown in Ref. \cite{Zeng_NLANE_2} that the non-linear anomalous Nernst coefficient has a dependence on the chemical potential similar to that of the non-linear anomalous Hall coefficient studied in the experiments of Ref.~\cite{kKang2019_NLAHE_experiment, qMa2019_NLAHE_experiment}. This is consistent with the analog of the Mott formula valid in the non-linear regime given in Eq.~(16).
To verify the relations between the coefficients of anomalous Hall and thermal Hall effects (namely the Wiedemann-Franz law in the non-linear regime given by Eq.~(14)), we compare the results for $l^{\prime}_y$ (index $z$ for $\bm{\Omega}_{\bm{k},z}$ is suppressed in a 2D system) based on Sommerfeld expansion in Eq.~(\ref{eq:NLTHE_Tdependence}) with that from  numerical calculations based on Eq.~(6).
As shown in Fig.~(3), the analytical results (red dotted line) from Eq.~(13, 14), coincide with the numerical results (the rest of the data besides the red dotted line) at low temperatures ($T=5 \sim 50K$). The numerical results and the prediction from the modified Wiedemann-Franz law differ from each other at higher temperatures ($T=100K \sim 300K$).  To verify the quadratic temperature dependence in Eq.~(13, 14), we plot the NLATHE coefficient $l^{\prime}_{y}$ as a function of $(k_B T)^2$ at different chemical potentials in Fig.~(\ref{fig:T_dependence}).
At low temperatures ($T\leq 50K$) the numerical (circles) and analytical (black lines) results are consistent with each other, while they start deviating from each other around $T=60K$ with $\mu=1.05 \Delta$ (blue circles). The deviations in Fig.~(3) and Fig.~(4) are due to the omission of the higher orders terms in temperature ($\mathcal{O}(T^4)$) in Eq.~(13). These contributions to $l^{\prime}_y$ can be ignored in the regime of low temperatures. We have checked that our results for NLATHE are robust against all the monotonous modulation of the band gap $\Delta$ such as tuning effect by external field \cite{Ashwin_2011_Efield}, finite temperature effects such as electron-phonon coupling \cite{Tongay_2012_T_gap}, doping effect through the mixing of chalcogens in MoX$_2$(X=S,Se,or Te) \cite{Ryder_2016_doping}, etc., as well as strength of the tilting parameter due to uni-axial strain~\cite{Habib_2015_strain, Riccardo_2017_strain,Lee_2019_strain,Aas_2018_strain}.

{\color{blue}{\em Discussion}}---
We derive the fundamental relations among the anomalous transport coefficients which replace the celebrated Wiedemann-Franz law and Mott relation in the non-linear regime. 
Important byproducts of these calculations include the prediction of non-linear anomalous thermal Hall effect (Eq. (5)) and the persistence of the non-linear anomalous Nernst coefficient in the zero temperature limit (Eq.~(15)). Our analytical results are confirmed by numerical calculations on MoS$_2$, a TR invariant TMDC that has been intensively studied in recent experiments. 
The non-linear Wiedemann-Franz law and Mott relation derived in this work are valid for topologically non-trivial conductors with non-zero BCD.

Along with the BCD-induced non-linear thermal Hall current $(j^Q_T)^{\prime}_a$ given in Eq.~(5), there exist other second-order contributions such as disorder-mediated contributions~\cite{Nandy_sysmetris_2019,zzDu2019_NLAHE_2} (non-linear side jump and skew-scattering contributions), scattering time independent contributions~\cite{tau0_niu,Nandy_sysmetris_2019} and Berry curvature independent contributions~\cite{Inti_2015_BCD}.
The BCD induced contributions to the non-linear response functions discussed in this work are dominant in TR invariant systems in which the Berry curvature independent contribution, which is non-zero only in the absence of both TRS and IS, discussed in Ref.~\cite{Naoto_NLANE_3} vanishes. Moreover, the Berry curvature induced contribution independent of scattering time which requires the breaking of TRS to be non-zero also vanishes in TR symmetric systems where the BCD induced contributions are dominant. In addition, experimentally, the external, disorder-mediated, side-jump and skew-scattering contributions to the non-linear response functions can be separated from the BCD induced contributions using a scaling formula as shown in Ref.~\cite{zzDu2019_NLAHE_2}. The Wiedemann-Franz law and Mott relation derived in this paper thus apply only to the BCD-induced anomalous part of the non-linear response functions which are non-zero in TR symmetric systems and leave out the contributions that take non-zero values only in systems with broken TRS.

{\color{blue}{\em Acknowledgments}}---C. Z. thank the useful discussions with Xiaoqin Yu. C. Z. and S. T. acknowledge support from ARO Grant
No. W911NF-16-1-0182.


\bibliography{my}

\begin{thebibliography}{55}%
\makeatletter
\providecommand \@ifxundefined [1]{%
 \@ifx{#1\undefined}
}%
\providecommand \@ifnum [1]{%
 \ifnum #1\expandafter \@firstoftwo
 \else \expandafter \@secondoftwo
 \fi
}%
\providecommand \@ifx [1]{%
 \ifx #1\expandafter \@firstoftwo
 \else \expandafter \@secondoftwo
 \fi
}%
\providecommand \natexlab [1]{#1}%
\providecommand \enquote  [1]{``#1''}%
\providecommand \bibnamefont  [1]{#1}%
\providecommand \bibfnamefont [1]{#1}%
\providecommand \citenamefont [1]{#1}%
\providecommand \href@noop [0]{\@secondoftwo}%
\providecommand \href [0]{\begingroup \@sanitize@url \@href}%
\providecommand \@href[1]{\@@startlink{#1}\@@href}%
\providecommand \@@href[1]{\endgroup#1\@@endlink}%
\providecommand \@sanitize@url [0]{\catcode `\\12\catcode `\$12\catcode
  `\&12\catcode `\#12\catcode `\^12\catcode `\_12\catcode `\%12\relax}%
\providecommand \@@startlink[1]{}%
\providecommand \@@endlink[0]{}%
\providecommand \url  [0]{\begingroup\@sanitize@url \@url }%
\providecommand \@url [1]{\endgroup\@href {#1}{\urlprefix }}%
\providecommand \urlprefix  [0]{URL }%
\providecommand \Eprint [0]{\href }%
\providecommand \doibase [0]{http://dx.doi.org/}%
\providecommand \selectlanguage [0]{\@gobble}%
\providecommand \bibinfo  [0]{\@secondoftwo}%
\providecommand \bibfield  [0]{\@secondoftwo}%
\providecommand \translation [1]{[#1]}%
\providecommand \BibitemOpen [0]{}%
\providecommand \bibitemStop [0]{}%
\providecommand \bibitemNoStop [0]{.\EOS\space}%
\providecommand \EOS [0]{\spacefactor3000\relax}%
\providecommand \BibitemShut  [1]{\csname bibitem#1\endcsname}%
\let\auto@bib@innerbib\@empty
\bibitem [{\citenamefont {Landau}\ and\ \citenamefont
  {Lifshitz}(1980)}]{LD_book}%
  \BibitemOpen
  \bibfield  {author} {\bibinfo {author} {\bibfnamefont {L.~D.}\ \bibnamefont
  {Landau}}\ and\ \bibinfo {author} {\bibfnamefont {E.~M.}\ \bibnamefont
  {Lifshitz}},\ }\href {\doibase
  https://doi.org/10.1016/B978-0-08-057046-4.50009-9} {\emph {\bibinfo {title}
  {Statistical Physics (Third Edition)}}}\ (\bibinfo  {publisher}
  {Butterworth-Heinemann},\ \bibinfo {address} {Oxford},\ \bibinfo {year}
  {1980})\BibitemShut {NoStop}%
\bibitem [{\citenamefont {Morimoto}\ and\ \citenamefont
  {Nagaosa}(2018)}]{Naoto_SP_2018}%
  \BibitemOpen
  \bibfield  {author} {\bibinfo {author} {\bibfnamefont {T.}~\bibnamefont
  {Morimoto}}\ and\ \bibinfo {author} {\bibfnamefont {N.}~\bibnamefont
  {Nagaosa}},\ }\href {\doibase 10.1038/s41598-018-20539-2} {\bibfield
  {journal} {\bibinfo  {journal} {Scientific Reports}\ }\textbf {\bibinfo
  {volume} {8}},\ \bibinfo {pages} {2045} (\bibinfo {year} {2018})}\BibitemShut
  {NoStop}%
\bibitem [{\citenamefont {Nagaosa}\ \emph {et~al.}(2010)\citenamefont
  {Nagaosa}, \citenamefont {Sinova}, \citenamefont {Onoda}, \citenamefont
  {MacDonald},\ and\ \citenamefont {Ong}}]{Naoto_2010review_berryphase}%
  \BibitemOpen
  \bibfield  {author} {\bibinfo {author} {\bibfnamefont {N.}~\bibnamefont
  {Nagaosa}}, \bibinfo {author} {\bibfnamefont {J.}~\bibnamefont {Sinova}},
  \bibinfo {author} {\bibfnamefont {S.}~\bibnamefont {Onoda}}, \bibinfo
  {author} {\bibfnamefont {A.~H.}\ \bibnamefont {MacDonald}}, \ and\ \bibinfo
  {author} {\bibfnamefont {N.~P.}\ \bibnamefont {Ong}},\ }\href {\doibase
  10.1103/RevModPhys.82.1539} {\bibfield  {journal} {\bibinfo  {journal} {Rev.
  Mod. Phys.}\ }\textbf {\bibinfo {volume} {82}},\ \bibinfo {pages} {1539}
  (\bibinfo {year} {2010})}\BibitemShut {NoStop}%
\bibitem [{\citenamefont {Xiao}\ \emph {et~al.}(2010)\citenamefont {Xiao},
  \citenamefont {Chang},\ and\ \citenamefont {Niu}}]{Niu_2010_berryphase}%
  \BibitemOpen
  \bibfield  {author} {\bibinfo {author} {\bibfnamefont {D.}~\bibnamefont
  {Xiao}}, \bibinfo {author} {\bibfnamefont {M.-C.}\ \bibnamefont {Chang}}, \
  and\ \bibinfo {author} {\bibfnamefont {Q.}~\bibnamefont {Niu}},\ }\href
  {\doibase 10.1103/RevModPhys.82.1959} {\bibfield  {journal} {\bibinfo
  {journal} {Rev. Mod. Phys.}\ }\textbf {\bibinfo {volume} {82}},\ \bibinfo
  {pages} {1959} (\bibinfo {year} {2010})}\BibitemShut {NoStop}%
\bibitem [{\citenamefont {Ye}\ \emph {et~al.}(1999)\citenamefont {Ye},
  \citenamefont {Kim}, \citenamefont {Millis}, \citenamefont {Shraiman},
  \citenamefont {Majumdar},\ and\ \citenamefont {Te\ifmmode \check{s}\else
  \v{s}\fi{}anovi\ifmmode~\acute{c}\else \'{c}\fi{}}}]{Ye_1999_berryphase}%
  \BibitemOpen
  \bibfield  {author} {\bibinfo {author} {\bibfnamefont {J.}~\bibnamefont
  {Ye}}, \bibinfo {author} {\bibfnamefont {Y.~B.}\ \bibnamefont {Kim}},
  \bibinfo {author} {\bibfnamefont {A.~J.}\ \bibnamefont {Millis}}, \bibinfo
  {author} {\bibfnamefont {B.~I.}\ \bibnamefont {Shraiman}}, \bibinfo {author}
  {\bibfnamefont {P.}~\bibnamefont {Majumdar}}, \ and\ \bibinfo {author}
  {\bibfnamefont {Z.}~\bibnamefont {Te\ifmmode \check{s}\else
  \v{s}\fi{}anovi\ifmmode~\acute{c}\else \'{c}\fi{}}},\ }\href {\doibase
  10.1103/PhysRevLett.83.3737} {\bibfield  {journal} {\bibinfo  {journal}
  {Phys. Rev. Lett.}\ }\textbf {\bibinfo {volume} {83}},\ \bibinfo {pages}
  {3737} (\bibinfo {year} {1999})}\BibitemShut {NoStop}%
\bibitem [{\citenamefont {Taguchi}\ \emph {et~al.}(2001)\citenamefont
  {Taguchi}, \citenamefont {Oohara}, \citenamefont {Yoshizawa}, \citenamefont
  {Nagaosa},\ and\ \citenamefont {Tokura}}]{Tokura2001_AHE}%
  \BibitemOpen
  \bibfield  {author} {\bibinfo {author} {\bibfnamefont {Y.}~\bibnamefont
  {Taguchi}}, \bibinfo {author} {\bibfnamefont {Y.}~\bibnamefont {Oohara}},
  \bibinfo {author} {\bibfnamefont {H.}~\bibnamefont {Yoshizawa}}, \bibinfo
  {author} {\bibfnamefont {N.}~\bibnamefont {Nagaosa}}, \ and\ \bibinfo
  {author} {\bibfnamefont {Y.}~\bibnamefont {Tokura}},\ }\href {\doibase
  10.1126/science.1058161} {\bibfield  {journal} {\bibinfo  {journal}
  {Science}\ }\textbf {\bibinfo {volume} {291}},\ \bibinfo {pages} {2573}
  (\bibinfo {year} {2001})}\BibitemShut {NoStop}%
\bibitem [{\citenamefont {Jungwirth}\ \emph {et~al.}(2002)\citenamefont
  {Jungwirth}, \citenamefont {Niu},\ and\ \citenamefont
  {MacDonald}}]{Niu2002_AHE}%
  \BibitemOpen
  \bibfield  {author} {\bibinfo {author} {\bibfnamefont {T.}~\bibnamefont
  {Jungwirth}}, \bibinfo {author} {\bibfnamefont {Q.}~\bibnamefont {Niu}}, \
  and\ \bibinfo {author} {\bibfnamefont {A.~H.}\ \bibnamefont {MacDonald}},\
  }\href {\doibase 10.1103/PhysRevLett.88.207208} {\bibfield  {journal}
  {\bibinfo  {journal} {Phys. Rev. Lett.}\ }\textbf {\bibinfo {volume} {88}},\
  \bibinfo {pages} {207208} (\bibinfo {year} {2002})}\BibitemShut {NoStop}%
\bibitem [{\citenamefont {Culcer}\ \emph {et~al.}(2003)\citenamefont {Culcer},
  \citenamefont {MacDonald},\ and\ \citenamefont {Niu}}]{Qian_2003_berryphase}%
  \BibitemOpen
  \bibfield  {author} {\bibinfo {author} {\bibfnamefont {D.}~\bibnamefont
  {Culcer}}, \bibinfo {author} {\bibfnamefont {A.}~\bibnamefont {MacDonald}}, \
  and\ \bibinfo {author} {\bibfnamefont {Q.}~\bibnamefont {Niu}},\ }\href
  {\doibase 10.1103/PhysRevB.68.045327} {\bibfield  {journal} {\bibinfo
  {journal} {Phys. Rev. B}\ }\textbf {\bibinfo {volume} {68}},\ \bibinfo
  {pages} {045327} (\bibinfo {year} {2003})}\BibitemShut {NoStop}%
\bibitem [{\citenamefont {Onoda}\ and\ \citenamefont
  {Nagaosa}(2003)}]{Nagaosa2003_AHE}%
  \BibitemOpen
  \bibfield  {author} {\bibinfo {author} {\bibfnamefont {M.}~\bibnamefont
  {Onoda}}\ and\ \bibinfo {author} {\bibfnamefont {N.}~\bibnamefont
  {Nagaosa}},\ }\href {\doibase 10.1103/PhysRevLett.90.206601} {\bibfield
  {journal} {\bibinfo  {journal} {Phys. Rev. Lett.}\ }\textbf {\bibinfo
  {volume} {90}},\ \bibinfo {pages} {206601} (\bibinfo {year}
  {2003})}\BibitemShut {NoStop}%
\bibitem [{\citenamefont {Lee}\ \emph {et~al.}(2004)\citenamefont {Lee},
  \citenamefont {Watauchi}, \citenamefont {Miller}, \citenamefont {Cava},\ and\
  \citenamefont {Ong}}]{Lee2004_AHE}%
  \BibitemOpen
  \bibfield  {author} {\bibinfo {author} {\bibfnamefont {W.-L.}\ \bibnamefont
  {Lee}}, \bibinfo {author} {\bibfnamefont {S.}~\bibnamefont {Watauchi}},
  \bibinfo {author} {\bibfnamefont {V.~L.}\ \bibnamefont {Miller}}, \bibinfo
  {author} {\bibfnamefont {R.~J.}\ \bibnamefont {Cava}}, \ and\ \bibinfo
  {author} {\bibfnamefont {N.~P.}\ \bibnamefont {Ong}},\ }\href {\doibase
  10.1126/science.1094383} {\bibfield  {journal} {\bibinfo  {journal}
  {Science}\ }\textbf {\bibinfo {volume} {303}},\ \bibinfo {pages} {1647}
  (\bibinfo {year} {2004})}\BibitemShut {NoStop}%
\bibitem [{\citenamefont {Fang}\ \emph {et~al.}(2003)\citenamefont {Fang},
  \citenamefont {Nagaosa}, \citenamefont {Takahashi}, \citenamefont {Asamitsu},
  \citenamefont {Mathieu}, \citenamefont {Ogasawara}, \citenamefont {Yamada},
  \citenamefont {Kawasaki}, \citenamefont {Tokura},\ and\ \citenamefont
  {Terakura}}]{Fang2003_AHE}%
  \BibitemOpen
  \bibfield  {author} {\bibinfo {author} {\bibfnamefont {Z.}~\bibnamefont
  {Fang}}, \bibinfo {author} {\bibfnamefont {N.}~\bibnamefont {Nagaosa}},
  \bibinfo {author} {\bibfnamefont {K.~S.}\ \bibnamefont {Takahashi}}, \bibinfo
  {author} {\bibfnamefont {A.}~\bibnamefont {Asamitsu}}, \bibinfo {author}
  {\bibfnamefont {R.}~\bibnamefont {Mathieu}}, \bibinfo {author} {\bibfnamefont
  {T.}~\bibnamefont {Ogasawara}}, \bibinfo {author} {\bibfnamefont
  {H.}~\bibnamefont {Yamada}}, \bibinfo {author} {\bibfnamefont
  {M.}~\bibnamefont {Kawasaki}}, \bibinfo {author} {\bibfnamefont
  {Y.}~\bibnamefont {Tokura}}, \ and\ \bibinfo {author} {\bibfnamefont
  {K.}~\bibnamefont {Terakura}},\ }\href {\doibase 10.1126/science.1089408}
  {\bibfield  {journal} {\bibinfo  {journal} {Science}\ }\textbf {\bibinfo
  {volume} {302}},\ \bibinfo {pages} {92} (\bibinfo {year} {2003})}\BibitemShut
  {NoStop}%
\bibitem [{\citenamefont {Haldane}(2004)}]{Haldane2004_AHE}%
  \BibitemOpen
  \bibfield  {author} {\bibinfo {author} {\bibfnamefont {F.~D.~M.}\
  \bibnamefont {Haldane}},\ }\href {\doibase 10.1103/PhysRevLett.93.206602}
  {\bibfield  {journal} {\bibinfo  {journal} {Phys. Rev. Lett.}\ }\textbf
  {\bibinfo {volume} {93}},\ \bibinfo {pages} {206602} (\bibinfo {year}
  {2004})}\BibitemShut {NoStop}%
\bibitem [{\citenamefont {Liu}\ \emph {et~al.}(2008)\citenamefont {Liu},
  \citenamefont {Qi}, \citenamefont {Dai}, \citenamefont {Fang},\ and\
  \citenamefont {Zhang}}]{cxliu_2008_breakingTRS}%
  \BibitemOpen
  \bibfield  {author} {\bibinfo {author} {\bibfnamefont {C.-X.}\ \bibnamefont
  {Liu}}, \bibinfo {author} {\bibfnamefont {X.-L.}\ \bibnamefont {Qi}},
  \bibinfo {author} {\bibfnamefont {X.}~\bibnamefont {Dai}}, \bibinfo {author}
  {\bibfnamefont {Z.}~\bibnamefont {Fang}}, \ and\ \bibinfo {author}
  {\bibfnamefont {S.-C.}\ \bibnamefont {Zhang}},\ }\href {\doibase
  10.1103/PhysRevLett.101.146802} {\bibfield  {journal} {\bibinfo  {journal}
  {Phys. Rev. Lett.}\ }\textbf {\bibinfo {volume} {101}},\ \bibinfo {pages}
  {146802} (\bibinfo {year} {2008})}\BibitemShut {NoStop}%
\bibitem [{\citenamefont {Yu}\ \emph {et~al.}(2010)\citenamefont {Yu},
  \citenamefont {Zhang}, \citenamefont {Zhang}, \citenamefont {Zhang},
  \citenamefont {Dai},\ and\ \citenamefont {Fang}}]{Yu2010_AHE}%
  \BibitemOpen
  \bibfield  {author} {\bibinfo {author} {\bibfnamefont {R.}~\bibnamefont
  {Yu}}, \bibinfo {author} {\bibfnamefont {W.}~\bibnamefont {Zhang}}, \bibinfo
  {author} {\bibfnamefont {H.-J.}\ \bibnamefont {Zhang}}, \bibinfo {author}
  {\bibfnamefont {S.-C.}\ \bibnamefont {Zhang}}, \bibinfo {author}
  {\bibfnamefont {X.}~\bibnamefont {Dai}}, \ and\ \bibinfo {author}
  {\bibfnamefont {Z.}~\bibnamefont {Fang}},\ }\href {\doibase
  10.1126/science.1187485} {\bibfield  {journal} {\bibinfo  {journal}
  {Science}\ }\textbf {\bibinfo {volume} {329}},\ \bibinfo {pages} {61}
  (\bibinfo {year} {2010})}\BibitemShut {NoStop}%
\bibitem [{\citenamefont {Chang}\ \emph {et~al.}(2013)\citenamefont {Chang},
  \citenamefont {Zhang}, \citenamefont {Feng}, \citenamefont {Shen},
  \citenamefont {Zhang}, \citenamefont {Guo}, \citenamefont {Li}, \citenamefont
  {Ou}, \citenamefont {Wei}, \citenamefont {Wang}, \citenamefont {Ji},
  \citenamefont {Feng}, \citenamefont {Ji}, \citenamefont {Chen}, \citenamefont
  {Jia}, \citenamefont {Dai}, \citenamefont {Fang}, \citenamefont {Zhang},
  \citenamefont {He}, \citenamefont {Wang}, \citenamefont {Lu}, \citenamefont
  {Ma},\ and\ \citenamefont {Xue}}]{qikun2013_AHE}%
  \BibitemOpen
  \bibfield  {author} {\bibinfo {author} {\bibfnamefont {C.-Z.}\ \bibnamefont
  {Chang}}, \bibinfo {author} {\bibfnamefont {J.}~\bibnamefont {Zhang}},
  \bibinfo {author} {\bibfnamefont {X.}~\bibnamefont {Feng}}, \bibinfo {author}
  {\bibfnamefont {J.}~\bibnamefont {Shen}}, \bibinfo {author} {\bibfnamefont
  {Z.}~\bibnamefont {Zhang}}, \bibinfo {author} {\bibfnamefont
  {M.}~\bibnamefont {Guo}}, \bibinfo {author} {\bibfnamefont {K.}~\bibnamefont
  {Li}}, \bibinfo {author} {\bibfnamefont {Y.}~\bibnamefont {Ou}}, \bibinfo
  {author} {\bibfnamefont {P.}~\bibnamefont {Wei}}, \bibinfo {author}
  {\bibfnamefont {L.-L.}\ \bibnamefont {Wang}}, \bibinfo {author}
  {\bibfnamefont {Z.-Q.}\ \bibnamefont {Ji}}, \bibinfo {author} {\bibfnamefont
  {Y.}~\bibnamefont {Feng}}, \bibinfo {author} {\bibfnamefont {S.}~\bibnamefont
  {Ji}}, \bibinfo {author} {\bibfnamefont {X.}~\bibnamefont {Chen}}, \bibinfo
  {author} {\bibfnamefont {J.}~\bibnamefont {Jia}}, \bibinfo {author}
  {\bibfnamefont {X.}~\bibnamefont {Dai}}, \bibinfo {author} {\bibfnamefont
  {Z.}~\bibnamefont {Fang}}, \bibinfo {author} {\bibfnamefont {S.-C.}\
  \bibnamefont {Zhang}}, \bibinfo {author} {\bibfnamefont {K.}~\bibnamefont
  {He}}, \bibinfo {author} {\bibfnamefont {Y.}~\bibnamefont {Wang}}, \bibinfo
  {author} {\bibfnamefont {L.}~\bibnamefont {Lu}}, \bibinfo {author}
  {\bibfnamefont {X.-C.}\ \bibnamefont {Ma}}, \ and\ \bibinfo {author}
  {\bibfnamefont {Q.-K.}\ \bibnamefont {Xue}},\ }\href {\doibase
  10.1126/science.1234414} {\bibfield  {journal} {\bibinfo  {journal}
  {Science}\ }\textbf {\bibinfo {volume} {340}},\ \bibinfo {pages} {167}
  (\bibinfo {year} {2013})}\BibitemShut {NoStop}%
\bibitem [{\citenamefont {Moore}\ and\ \citenamefont
  {Orenstein}(2010)}]{JMoore2010_Berryphase_factor}%
  \BibitemOpen
  \bibfield  {author} {\bibinfo {author} {\bibfnamefont {J.~E.}\ \bibnamefont
  {Moore}}\ and\ \bibinfo {author} {\bibfnamefont {J.}~\bibnamefont
  {Orenstein}},\ }\href {\doibase 10.1103/PhysRevLett.105.026805} {\bibfield
  {journal} {\bibinfo  {journal} {Phys. Rev. Lett.}\ }\textbf {\bibinfo
  {volume} {105}},\ \bibinfo {pages} {026805} (\bibinfo {year}
  {2010})}\BibitemShut {NoStop}%
\bibitem [{\citenamefont {Sodemann}\ and\ \citenamefont
  {Fu}(2015)}]{Inti_2015_BCD}%
  \BibitemOpen
  \bibfield  {author} {\bibinfo {author} {\bibfnamefont {I.}~\bibnamefont
  {Sodemann}}\ and\ \bibinfo {author} {\bibfnamefont {L.}~\bibnamefont {Fu}},\
  }\href {\doibase 10.1103/PhysRevLett.115.216806} {\bibfield  {journal}
  {\bibinfo  {journal} {Phys. Rev. Lett.}\ }\textbf {\bibinfo {volume} {115}},\
  \bibinfo {pages} {216806} (\bibinfo {year} {2015})}\BibitemShut {NoStop}%
\bibitem [{\citenamefont {Kang}\ \emph {et~al.}(2019)\citenamefont {Kang},
  \citenamefont {Li}, \citenamefont {Sohn}, \citenamefont {Shan},\ and\
  \citenamefont {Mak}}]{kKang2019_NLAHE_experiment}%
  \BibitemOpen
  \bibfield  {author} {\bibinfo {author} {\bibfnamefont {K.}~\bibnamefont
  {Kang}}, \bibinfo {author} {\bibfnamefont {T.}~\bibnamefont {Li}}, \bibinfo
  {author} {\bibfnamefont {E.}~\bibnamefont {Sohn}}, \bibinfo {author}
  {\bibfnamefont {J.}~\bibnamefont {Shan}}, \ and\ \bibinfo {author}
  {\bibfnamefont {K.~F.}\ \bibnamefont {Mak}},\ }\href {\doibase
  https://doi.org/10.1038/s41563-019-0294-7} {\bibfield  {journal} {\bibinfo
  {journal} {Nature Materials}\ }\textbf {\bibinfo {volume} {18}},\ \bibinfo
  {pages} {324} (\bibinfo {year} {2019})}\BibitemShut {NoStop}%
\bibitem [{\citenamefont {Ma}\ \emph {et~al.}(2019)\citenamefont {Ma},
  \citenamefont {Xu}, \citenamefont {Shen}, \citenamefont {MacNeill},
  \citenamefont {Fatemi}, \citenamefont {Chang}, \citenamefont {Valdivia},
  \citenamefont {Wu}, \citenamefont {Du}, \citenamefont {Hsu}, \citenamefont
  {Fang}, \citenamefont {Gibson}, \citenamefont {Watanabe}, \citenamefont
  {Taniguchi}, \citenamefont {Cava}, \citenamefont {Kaxiras}, \citenamefont
  {Lu}, \citenamefont {Lin}, \citenamefont {Fu}, \citenamefont {Gedik},\ and\
  \citenamefont {Jarillo-Herrero}}]{qMa2019_NLAHE_experiment}%
  \BibitemOpen
  \bibfield  {author} {\bibinfo {author} {\bibfnamefont {Q.}~\bibnamefont
  {Ma}}, \bibinfo {author} {\bibfnamefont {S.-Y.}\ \bibnamefont {Xu}}, \bibinfo
  {author} {\bibfnamefont {H.}~\bibnamefont {Shen}}, \bibinfo {author}
  {\bibfnamefont {D.}~\bibnamefont {MacNeill}}, \bibinfo {author}
  {\bibfnamefont {V.}~\bibnamefont {Fatemi}}, \bibinfo {author} {\bibfnamefont
  {T.-R.}\ \bibnamefont {Chang}}, \bibinfo {author} {\bibfnamefont {A.~M.~M.}\
  \bibnamefont {Valdivia}}, \bibinfo {author} {\bibfnamefont {S.}~\bibnamefont
  {Wu}}, \bibinfo {author} {\bibfnamefont {Z.}~\bibnamefont {Du}}, \bibinfo
  {author} {\bibfnamefont {C.-H.}\ \bibnamefont {Hsu}}, \bibinfo {author}
  {\bibfnamefont {S.}~\bibnamefont {Fang}}, \bibinfo {author} {\bibfnamefont
  {Q.~D.}\ \bibnamefont {Gibson}}, \bibinfo {author} {\bibfnamefont
  {K.}~\bibnamefont {Watanabe}}, \bibinfo {author} {\bibfnamefont
  {T.}~\bibnamefont {Taniguchi}}, \bibinfo {author} {\bibfnamefont {R.~J.}\
  \bibnamefont {Cava}}, \bibinfo {author} {\bibfnamefont {E.}~\bibnamefont
  {Kaxiras}}, \bibinfo {author} {\bibfnamefont {H.-Z.}\ \bibnamefont {Lu}},
  \bibinfo {author} {\bibfnamefont {H.}~\bibnamefont {Lin}}, \bibinfo {author}
  {\bibfnamefont {L.}~\bibnamefont {Fu}}, \bibinfo {author} {\bibfnamefont
  {N.}~\bibnamefont {Gedik}}, \ and\ \bibinfo {author} {\bibfnamefont
  {P.}~\bibnamefont {Jarillo-Herrero}},\ }\href {\doibase
  https://doi.org/10.1038/s41586-018-0807-6} {\bibfield  {journal} {\bibinfo
  {journal} {Nature}\ }\textbf {\bibinfo {volume} {565}},\ \bibinfo {pages}
  {337–342} (\bibinfo {year} {2019})}\BibitemShut {NoStop}%
\bibitem [{\citenamefont {Nandy}\ and\ \citenamefont
  {Sodemann}(2019)}]{Nandy_sysmetris_2019}%
  \BibitemOpen
  \bibfield  {author} {\bibinfo {author} {\bibfnamefont {S.}~\bibnamefont
  {Nandy}}\ and\ \bibinfo {author} {\bibfnamefont {I.}~\bibnamefont
  {Sodemann}},\ }\href {\doibase 10.1103/PhysRevB.100.195117} {\bibfield
  {journal} {\bibinfo  {journal} {Phys. Rev. B}\ }\textbf {\bibinfo {volume}
  {100}},\ \bibinfo {pages} {195117} (\bibinfo {year} {2019})}\BibitemShut
  {NoStop}%
\bibitem [{\citenamefont {Zhou}\ \emph {et~al.}(2019)\citenamefont {Zhou},
  \citenamefont {Zhang},\ and\ \citenamefont {Law}}]{Law2019_NLAHE_3}%
  \BibitemOpen
  \bibfield  {author} {\bibinfo {author} {\bibfnamefont {B.~T.}\ \bibnamefont
  {Zhou}}, \bibinfo {author} {\bibfnamefont {C.-P.}\ \bibnamefont {Zhang}}, \
  and\ \bibinfo {author} {\bibfnamefont {K.~T.}\ \bibnamefont {Law}},\ }\href
  {https://arxiv.org/abs/1903.11958} {\bibfield  {journal} {\bibinfo  {journal}
  {arXiv:1903.11958}\ } (\bibinfo {year} {2019})}\BibitemShut {NoStop}%
\bibitem [{\citenamefont {S\'anchez}\ and\ \citenamefont
  {B\"uttiker}(2004)}]{eeinteraction_1}%
  \BibitemOpen
  \bibfield  {author} {\bibinfo {author} {\bibfnamefont {D.}~\bibnamefont
  {S\'anchez}}\ and\ \bibinfo {author} {\bibfnamefont {M.}~\bibnamefont
  {B\"uttiker}},\ }\href {\doibase 10.1103/PhysRevLett.93.106802} {\bibfield
  {journal} {\bibinfo  {journal} {Phys. Rev. Lett.}\ }\textbf {\bibinfo
  {volume} {93}},\ \bibinfo {pages} {106802} (\bibinfo {year}
  {2004})}\BibitemShut {NoStop}%
\bibitem [{\citenamefont {Spivak}\ and\ \citenamefont
  {Zyuzin}(2004)}]{eeinteraction_2}%
  \BibitemOpen
  \bibfield  {author} {\bibinfo {author} {\bibfnamefont {B.}~\bibnamefont
  {Spivak}}\ and\ \bibinfo {author} {\bibfnamefont {A.}~\bibnamefont
  {Zyuzin}},\ }\href {\doibase 10.1103/PhysRevLett.93.226801} {\bibfield
  {journal} {\bibinfo  {journal} {Phys. Rev. Lett.}\ }\textbf {\bibinfo
  {volume} {93}},\ \bibinfo {pages} {226801} (\bibinfo {year}
  {2004})}\BibitemShut {NoStop}%
\bibitem [{\citenamefont {Nakai}\ and\ \citenamefont
  {Nagaosa}(2019)}]{Naoto_NLANE_3}%
  \BibitemOpen
  \bibfield  {author} {\bibinfo {author} {\bibfnamefont {R.}~\bibnamefont
  {Nakai}}\ and\ \bibinfo {author} {\bibfnamefont {N.}~\bibnamefont
  {Nagaosa}},\ }\href {\doibase 10.1103/PhysRevB.99.115201} {\bibfield
  {journal} {\bibinfo  {journal} {Phys. Rev. B}\ }\textbf {\bibinfo {volume}
  {99}},\ \bibinfo {pages} {115201} (\bibinfo {year} {2019})}\BibitemShut
  {NoStop}%
\bibitem [{\citenamefont {Yu}\ \emph {et~al.}(2019)\citenamefont {Yu},
  \citenamefont {Zhu}, \citenamefont {You}, \citenamefont {Low},\ and\
  \citenamefont {Su}}]{Yu_NLANE_1}%
  \BibitemOpen
  \bibfield  {author} {\bibinfo {author} {\bibfnamefont {X.-Q.}\ \bibnamefont
  {Yu}}, \bibinfo {author} {\bibfnamefont {Z.-G.}\ \bibnamefont {Zhu}},
  \bibinfo {author} {\bibfnamefont {J.-S.}\ \bibnamefont {You}}, \bibinfo
  {author} {\bibfnamefont {T.}~\bibnamefont {Low}}, \ and\ \bibinfo {author}
  {\bibfnamefont {G.}~\bibnamefont {Su}},\ }\href {\doibase
  10.1103/PhysRevB.99.201410} {\bibfield  {journal} {\bibinfo  {journal} {Phys.
  Rev. B}\ }\textbf {\bibinfo {volume} {99}},\ \bibinfo {pages} {201410}
  (\bibinfo {year} {2019})}\BibitemShut {NoStop}%
\bibitem [{\citenamefont {Zeng}\ \emph {et~al.}(2019)\citenamefont {Zeng},
  \citenamefont {Nandy}, \citenamefont {Taraphder},\ and\ \citenamefont
  {Tewari}}]{Zeng_NLANE_2}%
  \BibitemOpen
  \bibfield  {author} {\bibinfo {author} {\bibfnamefont {C.}~\bibnamefont
  {Zeng}}, \bibinfo {author} {\bibfnamefont {S.}~\bibnamefont {Nandy}},
  \bibinfo {author} {\bibfnamefont {A.}~\bibnamefont {Taraphder}}, \ and\
  \bibinfo {author} {\bibfnamefont {S.}~\bibnamefont {Tewari}},\ }\href
  {\doibase 10.1103/PhysRevB.100.245102} {\bibfield  {journal} {\bibinfo
  {journal} {Phys. Rev. B}\ }\textbf {\bibinfo {volume} {100}},\ \bibinfo
  {pages} {245102} (\bibinfo {year} {2019})}\BibitemShut {NoStop}%
\bibitem [{\citenamefont {Ashcroft}\ and\ \citenamefont
  {Mermin}(1976)}]{Ashcroft}%
  \BibitemOpen
  \bibfield  {author} {\bibinfo {author} {\bibfnamefont {N.~W.}\ \bibnamefont
  {Ashcroft}}\ and\ \bibinfo {author} {\bibfnamefont {j.~a.}\ \bibnamefont
  {Mermin}, \bibfnamefont {N.~David}},\ }\href
  {http://openlibrary.org/books/OL5048917M} {\emph {\bibinfo {title} {Solid
  state physics}}}\ (\bibinfo  {publisher} {New York Holt, Rinehart and
  Winston},\ \bibinfo {year} {1976})\ \bibinfo {note} {includes bibliographical
  references}\BibitemShut {NoStop}%
\bibitem [{\citenamefont {Xiao}\ \emph {et~al.}(2006)\citenamefont {Xiao},
  \citenamefont {Yao}, \citenamefont {Fang},\ and\ \citenamefont
  {Niu}}]{Xiao_2006_THE_wavepacket}%
  \BibitemOpen
  \bibfield  {author} {\bibinfo {author} {\bibfnamefont {D.}~\bibnamefont
  {Xiao}}, \bibinfo {author} {\bibfnamefont {Y.}~\bibnamefont {Yao}}, \bibinfo
  {author} {\bibfnamefont {Z.}~\bibnamefont {Fang}}, \ and\ \bibinfo {author}
  {\bibfnamefont {Q.}~\bibnamefont {Niu}},\ }\href {\doibase
  10.1103/PhysRevLett.97.026603} {\bibfield  {journal} {\bibinfo  {journal}
  {Phys. Rev. Lett.}\ }\textbf {\bibinfo {volume} {97}},\ \bibinfo {pages}
  {026603} (\bibinfo {year} {2006})}\BibitemShut {NoStop}%
\bibitem [{\citenamefont {{Qin}}\ \emph {et~al.}(2011)\citenamefont {{Qin}},
  \citenamefont {{Niu}},\ and\ \citenamefont
  {{Shi}}}]{QIN_2011_THE_wavepacket}%
  \BibitemOpen
  \bibfield  {author} {\bibinfo {author} {\bibfnamefont {T.}~\bibnamefont
  {{Qin}}}, \bibinfo {author} {\bibfnamefont {Q.}~\bibnamefont {{Niu}}}, \ and\
  \bibinfo {author} {\bibfnamefont {J.}~\bibnamefont {{Shi}}},\ }\href
  {\doibase 10.1103/PhysRevLett.107.236601} {\bibfield  {journal} {\bibinfo
  {journal} {\prl}\ }\textbf {\bibinfo {volume} {107}},\ \bibinfo {eid}
  {236601} (\bibinfo {year} {2011})},\ \Eprint {http://arxiv.org/abs/1108.3879}
  {arXiv:1108.3879 [cond-mat.stat-mech]} \BibitemShut {NoStop}%
\bibitem [{\citenamefont {{Zhang}}(2016)}]{lifa_2016_THE_wavepacket}%
  \BibitemOpen
  \bibfield  {author} {\bibinfo {author} {\bibfnamefont {L.}~\bibnamefont
  {{Zhang}}},\ }\href {\doibase 10.1088/1367-2630/18/10/103039} {\bibfield
  {journal} {\bibinfo  {journal} {New Journal of Physics}\ }\textbf {\bibinfo
  {volume} {18}},\ \bibinfo {eid} {103039} (\bibinfo {year}
  {2016})}\BibitemShut {NoStop}%
\bibitem [{\citenamefont {{Ferreiros}}\ \emph {et~al.}(2017)\citenamefont
  {{Ferreiros}}, \citenamefont {{Zyuzin}},\ and\ \citenamefont
  {{Bardarson}}}]{Jens_2017_THE}%
  \BibitemOpen
  \bibfield  {author} {\bibinfo {author} {\bibfnamefont {Y.}~\bibnamefont
  {{Ferreiros}}}, \bibinfo {author} {\bibfnamefont {A.~A.}\ \bibnamefont
  {{Zyuzin}}}, \ and\ \bibinfo {author} {\bibfnamefont {J.~H.}\ \bibnamefont
  {{Bardarson}}},\ }\href {\doibase 10.1103/PhysRevB.96.115202} {\bibfield
  {journal} {\bibinfo  {journal} {\prb}\ }\textbf {\bibinfo {volume} {96}},\
  \bibinfo {eid} {115202} (\bibinfo {year} {2017})},\ \Eprint
  {http://arxiv.org/abs/1707.01444} {arXiv:1707.01444 [cond-mat.mes-hall]}
  \BibitemShut {NoStop}%
\bibitem [{\citenamefont {Lundgren}\ \emph {et~al.}(2014)\citenamefont
  {Lundgren}, \citenamefont {Laurell},\ and\ \citenamefont
  {Fiete}}]{Lundgren_2014_THE}%
  \BibitemOpen
  \bibfield  {author} {\bibinfo {author} {\bibfnamefont {R.}~\bibnamefont
  {Lundgren}}, \bibinfo {author} {\bibfnamefont {P.}~\bibnamefont {Laurell}}, \
  and\ \bibinfo {author} {\bibfnamefont {G.~A.}\ \bibnamefont {Fiete}},\ }\href
  {\doibase 10.1103/PhysRevB.90.165115} {\bibfield  {journal} {\bibinfo
  {journal} {Phys. Rev. B}\ }\textbf {\bibinfo {volume} {90}},\ \bibinfo
  {pages} {165115} (\bibinfo {year} {2014})}\BibitemShut {NoStop}%
\bibitem [{\citenamefont {Yokoyama}\ and\ \citenamefont
  {Murakami}(2011)}]{Yokoyama_2011_THE_heat}%
  \BibitemOpen
  \bibfield  {author} {\bibinfo {author} {\bibfnamefont {T.}~\bibnamefont
  {Yokoyama}}\ and\ \bibinfo {author} {\bibfnamefont {S.}~\bibnamefont
  {Murakami}},\ }\href {\doibase 10.1103/PhysRevB.83.161407} {\bibfield
  {journal} {\bibinfo  {journal} {Phys. Rev. B}\ }\textbf {\bibinfo {volume}
  {83}},\ \bibinfo {pages} {161407} (\bibinfo {year} {2011})}\BibitemShut
  {NoStop}%
\bibitem [{\citenamefont {Sharma}\ \emph {et~al.}(2016)\citenamefont {Sharma},
  \citenamefont {Goswami},\ and\ \citenamefont {Tewari}}]{Girish_2016_THE}%
  \BibitemOpen
  \bibfield  {author} {\bibinfo {author} {\bibfnamefont {G.}~\bibnamefont
  {Sharma}}, \bibinfo {author} {\bibfnamefont {P.}~\bibnamefont {Goswami}}, \
  and\ \bibinfo {author} {\bibfnamefont {S.}~\bibnamefont {Tewari}},\ }\href
  {\doibase 10.1103/PhysRevB.93.035116} {\bibfield  {journal} {\bibinfo
  {journal} {Phys. Rev. B}\ }\textbf {\bibinfo {volume} {93}},\ \bibinfo
  {pages} {035116} (\bibinfo {year} {2016})}\BibitemShut {NoStop}%
\bibitem [{\citenamefont {McCormick}\ \emph {et~al.}(2017)\citenamefont
  {McCormick}, \citenamefont {McKay},\ and\ \citenamefont
  {Trivedi}}]{Timothy_2017_THE}%
  \BibitemOpen
  \bibfield  {author} {\bibinfo {author} {\bibfnamefont {T.~M.}\ \bibnamefont
  {McCormick}}, \bibinfo {author} {\bibfnamefont {R.~C.}\ \bibnamefont
  {McKay}}, \ and\ \bibinfo {author} {\bibfnamefont {N.}~\bibnamefont
  {Trivedi}},\ }\href {\doibase 10.1103/PhysRevB.96.235116} {\bibfield
  {journal} {\bibinfo  {journal} {Phys. Rev. B}\ }\textbf {\bibinfo {volume}
  {96}},\ \bibinfo {pages} {235116} (\bibinfo {year} {2017})}\BibitemShut
  {NoStop}%
\bibitem [{\citenamefont {Bergman}\ and\ \citenamefont
  {Oganesyan}(2010)}]{Doron_2010}%
  \BibitemOpen
  \bibfield  {author} {\bibinfo {author} {\bibfnamefont {D.~L.}\ \bibnamefont
  {Bergman}}\ and\ \bibinfo {author} {\bibfnamefont {V.}~\bibnamefont
  {Oganesyan}},\ }\href {\doibase 10.1103/PhysRevLett.104.066601} {\bibfield
  {journal} {\bibinfo  {journal} {Phys. Rev. Lett.}\ }\textbf {\bibinfo
  {volume} {104}},\ \bibinfo {pages} {066601} (\bibinfo {year}
  {2010})}\BibitemShut {NoStop}%
\bibitem [{\citenamefont {Franz}\ and\ \citenamefont
  {Wiedemann}(1853)}]{Lorenz_law}%
  \BibitemOpen
  \bibfield  {author} {\bibinfo {author} {\bibfnamefont {R.}~\bibnamefont
  {Franz}}\ and\ \bibinfo {author} {\bibfnamefont {G.}~\bibnamefont
  {Wiedemann}},\ }\href {\doibase 10.1002/andp.18531650802} {\bibfield
  {journal} {\bibinfo  {journal} {Annalen der Physik}\ }\textbf {\bibinfo
  {volume} {165}},\ \bibinfo {pages} {497} (\bibinfo {year}
  {1853})}\BibitemShut {NoStop}%
\bibitem [{\citenamefont {{L{\'o}pez}}\ and\ \citenamefont
  {{S{\'a}nchez}}(2013)}]{Rosa_2013_wfLaw_violation_NLTHE}%
  \BibitemOpen
  \bibfield  {author} {\bibinfo {author} {\bibfnamefont {R.}~\bibnamefont
  {{L{\'o}pez}}}\ and\ \bibinfo {author} {\bibfnamefont {D.}~\bibnamefont
  {{S{\'a}nchez}}},\ }\href {\doibase 10.1103/PhysRevB.88.045129} {\bibfield
  {journal} {\bibinfo  {journal} {\prb}\ }\textbf {\bibinfo {volume} {88}},\
  \bibinfo {eid} {045129} (\bibinfo {year} {2013})},\ \Eprint
  {http://arxiv.org/abs/1302.5557} {arXiv:1302.5557 [cond-mat.mes-hall]}
  \BibitemShut {NoStop}%
\bibitem [{\citenamefont {Kim}(2014)}]{violation_WF_kim}%
  \BibitemOpen
  \bibfield  {author} {\bibinfo {author} {\bibfnamefont {K.-S.}\ \bibnamefont
  {Kim}},\ }\href {\doibase 10.1103/PhysRevB.90.121108} {\bibfield  {journal}
  {\bibinfo  {journal} {Phys. Rev. B}\ }\textbf {\bibinfo {volume} {90}},\
  \bibinfo {pages} {121108} (\bibinfo {year} {2014})}\BibitemShut {NoStop}%
\bibitem [{\citenamefont {{L{\'o}pez}}\ \emph {et~al.}(2014)\citenamefont
  {{L{\'o}pez}}, \citenamefont {{Hwang}},\ and\ \citenamefont
  {{S{\'a}nchez}}}]{Rosa_2014_wfLaw_violation}%
  \BibitemOpen
  \bibfield  {author} {\bibinfo {author} {\bibfnamefont {R.}~\bibnamefont
  {{L{\'o}pez}}}, \bibinfo {author} {\bibfnamefont {S.-Y.}\ \bibnamefont
  {{Hwang}}}, \ and\ \bibinfo {author} {\bibfnamefont {D.}~\bibnamefont
  {{S{\'a}nchez}}},\ }in\ \href {\doibase 10.1088/1742-6596/568/5/052016}
  {\emph {\bibinfo {booktitle} {Journal of Physics Conference Series}}},\
  \bibinfo {series} {Journal of Physics Conference Series}, Vol.\ \bibinfo
  {volume} {568}\ (\bibinfo {year} {2014})\ p.\ \bibinfo {pages} {052016},\
  \Eprint {http://arxiv.org/abs/1505.03483} {arXiv:1505.03483
  [cond-mat.mes-hall]} \BibitemShut {NoStop}%
\bibitem [{\citenamefont {{Xu}}\ \emph {et~al.}(2018)\citenamefont {{Xu}},
  \citenamefont {{Li}}, \citenamefont {{Lu}}, \citenamefont {{Collignon}},
  \citenamefont {{Fu}}, \citenamefont {{Koo}}, \citenamefont {{Fauqu{\'e}}},
  \citenamefont {{Yan}}, \citenamefont {{Zhu}},\ and\ \citenamefont
  {{Behnia}}}]{Xu_wfLaw_2018_violation}%
  \BibitemOpen
  \bibfield  {author} {\bibinfo {author} {\bibfnamefont {L.}~\bibnamefont
  {{Xu}}}, \bibinfo {author} {\bibfnamefont {X.}~\bibnamefont {{Li}}}, \bibinfo
  {author} {\bibfnamefont {X.}~\bibnamefont {{Lu}}}, \bibinfo {author}
  {\bibfnamefont {C.}~\bibnamefont {{Collignon}}}, \bibinfo {author}
  {\bibfnamefont {H.}~\bibnamefont {{Fu}}}, \bibinfo {author} {\bibfnamefont
  {J.}~\bibnamefont {{Koo}}}, \bibinfo {author} {\bibfnamefont
  {B.}~\bibnamefont {{Fauqu{\'e}}}}, \bibinfo {author} {\bibfnamefont
  {B.}~\bibnamefont {{Yan}}}, \bibinfo {author} {\bibfnamefont
  {Z.}~\bibnamefont {{Zhu}}}, \ and\ \bibinfo {author} {\bibfnamefont
  {K.}~\bibnamefont {{Behnia}}},\ }\href@noop {} {\bibfield  {journal}
  {\bibinfo  {journal} {arXiv e-prints}\ ,\ \bibinfo {eid} {arXiv:1812.04339}}
  (\bibinfo {year} {2018})},\ \Eprint {http://arxiv.org/abs/1812.04339}
  {arXiv:1812.04339 [cond-mat.str-el]} \BibitemShut {NoStop}%
\bibitem [{\citenamefont {{Jaoui}}\ \emph {et~al.}(2018)\citenamefont
  {{Jaoui}}, \citenamefont {{Fauqu{\'e}}}, \citenamefont {{Rischau}},
  \citenamefont {{Subedi}}, \citenamefont {{Fu}}, \citenamefont {{Gooth}},
  \citenamefont {{Kumar}}, \citenamefont {{S{\"u}{\ss}}}, \citenamefont
  {{Maslov}}, \citenamefont {{Felser}},\ and\ \citenamefont
  {{Behnia}}}]{Jaoui_2018_wfLaw_violation}%
  \BibitemOpen
  \bibfield  {author} {\bibinfo {author} {\bibfnamefont {A.}~\bibnamefont
  {{Jaoui}}}, \bibinfo {author} {\bibfnamefont {B.}~\bibnamefont
  {{Fauqu{\'e}}}}, \bibinfo {author} {\bibfnamefont {C.~W.}\ \bibnamefont
  {{Rischau}}}, \bibinfo {author} {\bibfnamefont {A.}~\bibnamefont {{Subedi}}},
  \bibinfo {author} {\bibfnamefont {C.}~\bibnamefont {{Fu}}}, \bibinfo {author}
  {\bibfnamefont {J.}~\bibnamefont {{Gooth}}}, \bibinfo {author} {\bibfnamefont
  {N.}~\bibnamefont {{Kumar}}}, \bibinfo {author} {\bibfnamefont
  {V.}~\bibnamefont {{S{\"u}{\ss}}}}, \bibinfo {author} {\bibfnamefont {D.~L.}\
  \bibnamefont {{Maslov}}}, \bibinfo {author} {\bibfnamefont {C.}~\bibnamefont
  {{Felser}}}, \ and\ \bibinfo {author} {\bibfnamefont {K.}~\bibnamefont
  {{Behnia}}},\ }\href {\doibase 10.1038/s41535-018-0136-x} {\bibfield
  {journal} {\bibinfo  {journal} {npj Quantum Materials}\ }\textbf {\bibinfo
  {volume} {3}},\ \bibinfo {eid} {64} (\bibinfo {year} {2018})},\ \Eprint
  {http://arxiv.org/abs/1806.04094} {arXiv:1806.04094 [cond-mat.str-el]}
  \BibitemShut {NoStop}%
\bibitem [{\citenamefont {Vinkler-Aviv}(2019)}]{Yuval_2019_wfLaw_wiolation}%
  \BibitemOpen
  \bibfield  {author} {\bibinfo {author} {\bibfnamefont {Y.}~\bibnamefont
  {Vinkler-Aviv}},\ }\href {\doibase 10.1103/PhysRevB.100.041106} {\bibfield
  {journal} {\bibinfo  {journal} {Phys. Rev. B}\ }\textbf {\bibinfo {volume}
  {100}},\ \bibinfo {pages} {041106} (\bibinfo {year} {2019})}\BibitemShut
  {NoStop}%
\bibitem [{\citenamefont {Nandy}\ \emph {et~al.}(2019)\citenamefont {Nandy},
  \citenamefont {Taraphder},\ and\ \citenamefont
  {Tewari}}]{Nandy_2019_violation}%
  \BibitemOpen
  \bibfield  {author} {\bibinfo {author} {\bibfnamefont {S.}~\bibnamefont
  {Nandy}}, \bibinfo {author} {\bibfnamefont {A.}~\bibnamefont {Taraphder}}, \
  and\ \bibinfo {author} {\bibfnamefont {S.}~\bibnamefont {Tewari}},\ }\href
  {\doibase 10.1103/PhysRevB.100.115139} {\bibfield  {journal} {\bibinfo
  {journal} {Phys. Rev. B}\ }\textbf {\bibinfo {volume} {100}},\ \bibinfo
  {pages} {115139} (\bibinfo {year} {2019})}\BibitemShut {NoStop}%
\bibitem [{\citenamefont {{Manzeli}}\ \emph {et~al.}(2017)\citenamefont
  {{Manzeli}}, \citenamefont {{Ovchinnikov}}, \citenamefont {{Pasquier}},
  \citenamefont {{Yazyev}},\ and\ \citenamefont {{Kis}}}]{Sajedeh_2017_TMDCs}%
  \BibitemOpen
  \bibfield  {author} {\bibinfo {author} {\bibfnamefont {S.}~\bibnamefont
  {{Manzeli}}}, \bibinfo {author} {\bibfnamefont {D.}~\bibnamefont
  {{Ovchinnikov}}}, \bibinfo {author} {\bibfnamefont {D.}~\bibnamefont
  {{Pasquier}}}, \bibinfo {author} {\bibfnamefont {O.~V.}\ \bibnamefont
  {{Yazyev}}}, \ and\ \bibinfo {author} {\bibfnamefont {A.}~\bibnamefont
  {{Kis}}},\ }\href {\doibase 10.1038/natrevmats.2017.33} {\bibfield  {journal}
  {\bibinfo  {journal} {Nature Reviews Materials}\ }\textbf {\bibinfo {volume}
  {2}},\ \bibinfo {pages} {17033} (\bibinfo {year} {2017})}\BibitemShut
  {NoStop}%
\bibitem [{\citenamefont {{Mak}}\ and\ \citenamefont
  {{Shan}}(2016)}]{Kin_2016_TMDCs}%
  \BibitemOpen
  \bibfield  {author} {\bibinfo {author} {\bibfnamefont {K.~F.}\ \bibnamefont
  {{Mak}}}\ and\ \bibinfo {author} {\bibfnamefont {J.}~\bibnamefont {{Shan}}},\
  }\href {\doibase 10.1038/nphoton.2015.282} {\bibfield  {journal} {\bibinfo
  {journal} {Nature Photonics}\ }\textbf {\bibinfo {volume} {10}},\ \bibinfo
  {pages} {216} (\bibinfo {year} {2016})}\BibitemShut {NoStop}%
\bibitem [{\citenamefont {Ramasubramaniam}\ \emph {et~al.}(2011)\citenamefont
  {Ramasubramaniam}, \citenamefont {Naveh},\ and\ \citenamefont
  {Towe}}]{Ashwin_2011_Efield}%
  \BibitemOpen
  \bibfield  {author} {\bibinfo {author} {\bibfnamefont {A.}~\bibnamefont
  {Ramasubramaniam}}, \bibinfo {author} {\bibfnamefont {D.}~\bibnamefont
  {Naveh}}, \ and\ \bibinfo {author} {\bibfnamefont {E.}~\bibnamefont {Towe}},\
  }\href {\doibase 10.1103/PhysRevB.84.205325} {\bibfield  {journal} {\bibinfo
  {journal} {Phys. Rev. B}\ }\textbf {\bibinfo {volume} {84}},\ \bibinfo
  {pages} {205325} (\bibinfo {year} {2011})}\BibitemShut {NoStop}%
\bibitem [{\citenamefont {Tongay}\ \emph {et~al.}(2012)\citenamefont {Tongay},
  \citenamefont {Zhou}, \citenamefont {Ataca}, \citenamefont {Lo},
  \citenamefont {Matthews}, \citenamefont {Li}, \citenamefont {Grossman},\ and\
  \citenamefont {Wu}}]{Tongay_2012_T_gap}%
  \BibitemOpen
  \bibfield  {author} {\bibinfo {author} {\bibfnamefont {S.}~\bibnamefont
  {Tongay}}, \bibinfo {author} {\bibfnamefont {J.}~\bibnamefont {Zhou}},
  \bibinfo {author} {\bibfnamefont {C.}~\bibnamefont {Ataca}}, \bibinfo
  {author} {\bibfnamefont {K.}~\bibnamefont {Lo}}, \bibinfo {author}
  {\bibfnamefont {T.~S.}\ \bibnamefont {Matthews}}, \bibinfo {author}
  {\bibfnamefont {J.}~\bibnamefont {Li}}, \bibinfo {author} {\bibfnamefont
  {J.~C.}\ \bibnamefont {Grossman}}, \ and\ \bibinfo {author} {\bibfnamefont
  {J.}~\bibnamefont {Wu}},\ }\href {\doibase 10.1021/nl302584w} {\bibfield
  {journal} {\bibinfo  {journal} {Nano Letters}\ }\textbf {\bibinfo {volume}
  {12}},\ \bibinfo {pages} {5576} (\bibinfo {year} {2012})},\ \bibinfo {note}
  {pMID: 23098085}\BibitemShut {NoStop}%
\bibitem [{\citenamefont {{Ryder}}\ \emph {et~al.}(2016)\citenamefont
  {{Ryder}}, \citenamefont {{Wood}}, \citenamefont {{Wells}},\ and\
  \citenamefont {{Hersam}}}]{Ryder_2016_doping}%
  \BibitemOpen
  \bibfield  {author} {\bibinfo {author} {\bibfnamefont {C.~R.}\ \bibnamefont
  {{Ryder}}}, \bibinfo {author} {\bibfnamefont {J.~D.}\ \bibnamefont {{Wood}}},
  \bibinfo {author} {\bibfnamefont {S.~A.}\ \bibnamefont {{Wells}}}, \ and\
  \bibinfo {author} {\bibfnamefont {M.~C.}\ \bibnamefont {{Hersam}}},\
  }\href@noop {} {\bibfield  {journal} {\bibinfo  {journal} {arXiv e-prints}\
  ,\ \bibinfo {eid} {arXiv:1603.08544}} (\bibinfo {year} {2016})},\ \Eprint
  {http://arxiv.org/abs/1603.08544} {arXiv:1603.08544 [cond-mat.mtrl-sci]}
  \BibitemShut {NoStop}%
\bibitem [{\citenamefont {Rostami}\ \emph {et~al.}(2015)\citenamefont
  {Rostami}, \citenamefont {Rold\'an}, \citenamefont {Cappelluti},
  \citenamefont {Asgari},\ and\ \citenamefont {Guinea}}]{Habib_2015_strain}%
  \BibitemOpen
  \bibfield  {author} {\bibinfo {author} {\bibfnamefont {H.}~\bibnamefont
  {Rostami}}, \bibinfo {author} {\bibfnamefont {R.}~\bibnamefont {Rold\'an}},
  \bibinfo {author} {\bibfnamefont {E.}~\bibnamefont {Cappelluti}}, \bibinfo
  {author} {\bibfnamefont {R.}~\bibnamefont {Asgari}}, \ and\ \bibinfo {author}
  {\bibfnamefont {F.}~\bibnamefont {Guinea}},\ }\href {\doibase
  10.1103/PhysRevB.92.195402} {\bibfield  {journal} {\bibinfo  {journal} {Phys.
  Rev. B}\ }\textbf {\bibinfo {volume} {92}},\ \bibinfo {pages} {195402}
  (\bibinfo {year} {2015})}\BibitemShut {NoStop}%
\bibitem [{\citenamefont {{Frisenda}}\ \emph {et~al.}(2017)\citenamefont
  {{Frisenda}}, \citenamefont {{Dr{\"u}ppel}}, \citenamefont {{Schmidt}},
  \citenamefont {{Michaelis de Vasconcellos}}, \citenamefont {{Perez de Lara}},
  \citenamefont {{Bratschitsch}}, \citenamefont {{Rohlfing}},\ and\
  \citenamefont {{Castellanos-Gomez}}}]{Riccardo_2017_strain}%
  \BibitemOpen
  \bibfield  {author} {\bibinfo {author} {\bibfnamefont {R.}~\bibnamefont
  {{Frisenda}}}, \bibinfo {author} {\bibfnamefont {M.}~\bibnamefont
  {{Dr{\"u}ppel}}}, \bibinfo {author} {\bibfnamefont {R.}~\bibnamefont
  {{Schmidt}}}, \bibinfo {author} {\bibfnamefont {S.}~\bibnamefont {{Michaelis
  de Vasconcellos}}}, \bibinfo {author} {\bibfnamefont {D.}~\bibnamefont
  {{Perez de Lara}}}, \bibinfo {author} {\bibfnamefont {R.}~\bibnamefont
  {{Bratschitsch}}}, \bibinfo {author} {\bibfnamefont {M.}~\bibnamefont
  {{Rohlfing}}}, \ and\ \bibinfo {author} {\bibfnamefont {A.}~\bibnamefont
  {{Castellanos-Gomez}}},\ }\href@noop {} {\bibfield  {journal} {\bibinfo
  {journal} {arXiv e-prints}\ ,\ \bibinfo {eid} {arXiv:1703.02831}} (\bibinfo
  {year} {2017})},\ \Eprint {http://arxiv.org/abs/1703.02831} {arXiv:1703.02831
  [cond-mat.mes-hall]} \BibitemShut {NoStop}%
\bibitem [{\citenamefont {Son}\ \emph {et~al.}(2019)\citenamefont {Son},
  \citenamefont {Kim}, \citenamefont {Ahn}, \citenamefont {Lee},\ and\
  \citenamefont {Lee}}]{Lee_2019_strain}%
  \BibitemOpen
  \bibfield  {author} {\bibinfo {author} {\bibfnamefont {J.}~\bibnamefont
  {Son}}, \bibinfo {author} {\bibfnamefont {K.-H.}\ \bibnamefont {Kim}},
  \bibinfo {author} {\bibfnamefont {Y.~H.}\ \bibnamefont {Ahn}}, \bibinfo
  {author} {\bibfnamefont {H.-W.}\ \bibnamefont {Lee}}, \ and\ \bibinfo
  {author} {\bibfnamefont {J.}~\bibnamefont {Lee}},\ }\href {\doibase
  10.1103/PhysRevLett.123.036806} {\bibfield  {journal} {\bibinfo  {journal}
  {Phys. Rev. Lett.}\ }\textbf {\bibinfo {volume} {123}},\ \bibinfo {pages}
  {036806} (\bibinfo {year} {2019})}\BibitemShut {NoStop}%
\bibitem [{\citenamefont {{Aas}}\ and\ \citenamefont
  {{Bulutay}}(2018)}]{Aas_2018_strain}%
  \BibitemOpen
  \bibfield  {author} {\bibinfo {author} {\bibfnamefont {S.}~\bibnamefont
  {{Aas}}}\ and\ \bibinfo {author} {\bibfnamefont {C.}~\bibnamefont
  {{Bulutay}}},\ }\href {\doibase 10.1364/OE.26.028672} {\bibfield  {journal}
  {\bibinfo  {journal} {Optics Express}\ }\textbf {\bibinfo {volume} {26}},\
  \bibinfo {pages} {28672} (\bibinfo {year} {2018})},\ \Eprint
  {http://arxiv.org/abs/1807.08568} {arXiv:1807.08568 [cond-mat.mes-hall]}
  \BibitemShut {NoStop}%
\bibitem [{\citenamefont {{Du}}\ \emph {et~al.}(2019)\citenamefont {{Du}},
  \citenamefont {{Wang}}, \citenamefont {{Li}}, \citenamefont {{Lu}},\ and\
  \citenamefont {{Xie}}}]{zzDu2019_NLAHE_2}%
  \BibitemOpen
  \bibfield  {author} {\bibinfo {author} {\bibfnamefont {Z.~Z.}\ \bibnamefont
  {{Du}}}, \bibinfo {author} {\bibfnamefont {C.~M.}\ \bibnamefont {{Wang}}},
  \bibinfo {author} {\bibfnamefont {S.}~\bibnamefont {{Li}}}, \bibinfo {author}
  {\bibfnamefont {H.-Z.}\ \bibnamefont {{Lu}}}, \ and\ \bibinfo {author}
  {\bibfnamefont {X.~C.}\ \bibnamefont {{Xie}}},\ }\href {\doibase
  10.1038/s41467-019-10941-3} {\bibfield  {journal} {\bibinfo  {journal}
  {Nature Communications}\ }\textbf {\bibinfo {volume} {10}},\ \bibinfo {eid}
  {3047} (\bibinfo {year} {2019})},\ \Eprint {http://arxiv.org/abs/1812.08377}
  {arXiv:1812.08377 [cond-mat.mes-hall]} \BibitemShut {NoStop}%
\bibitem [{\citenamefont {Gao}\ \emph {et~al.}(2014)\citenamefont {Gao},
  \citenamefont {Yang},\ and\ \citenamefont {Niu}}]{tau0_niu}%
  \BibitemOpen
  \bibfield  {author} {\bibinfo {author} {\bibfnamefont {Y.}~\bibnamefont
  {Gao}}, \bibinfo {author} {\bibfnamefont {S.~A.}\ \bibnamefont {Yang}}, \
  and\ \bibinfo {author} {\bibfnamefont {Q.}~\bibnamefont {Niu}},\ }\href
  {\doibase 10.1103/PhysRevLett.112.166601} {\bibfield  {journal} {\bibinfo
  {journal} {Phys. Rev. Lett.}\ }\textbf {\bibinfo {volume} {112}},\ \bibinfo
  {pages} {166601} (\bibinfo {year} {2014})}\BibitemShut {NoStop}%
\end{thebibliography}%


\end{document}